# When does lasing become a condensation phenomenon?*

**Baruch Fischer* and Rafi Weill**

*Department of Electrical Engineering, Technion, Haifa 32000, Israel*

*fischer@ee.technion.ac.il*

**Abstract:**

We present a generic route to classical light condensation (LC) in linear photonic mode systems, such as cw lasers, with different grounds from regular Bose-Einstein condensation (BEC). LC is based on weighting the modes in a noisy environment (spontaneous emission, etc.) in a loss-gain scale, rather than in photon energy. It is characterized by a sharp transition from a multi- to single-mode oscillation. The study uses a linear multivariate Langevin formulation which gives a mode occupation hierarchy that functions like Bose-Einstein statistics. Condensation occurs when the spectral filtering has near the lowest-loss mode a power law dependence with exponent smaller than 1. We then discuss how condensation can occur in photon systems, its relation to lasing and the difficulties to observe regular photon-BEC in laser cavities. We raise the possibility that experiments on photon condensation in optical cavities fall in a classical LC or lasing category rather than being a thermal-quantum BEC phenomenon.

---

\* This manuscript with a few changes was sent to Nature in June 2011 (along with a shorter BCA submission with critique on a Nature paper on photon-BEC observation [3]). It was afterwards sent to Nature Photonics in March 2012, and again in June 2012, but it was not published there. It presumably didn't meet some (quality or popular based /arbitrarily/hidden?) pre-selection-rules. Will anybody take the lead to challenge that system?

# 1. Introduction

Bose-Einstein condensation (BEC) was predicted in 1924-5, but was experimentally observed seventy years later with atomic particles at ultra-low temperatures [1,2]. The interesting question if and how can BEC occur with non-atomic bosons, such as photons, also attracted attention. A recent paper reported on the observation of photon-BEC in an optical micro-cavity [3,4] at a room temperature, and earlier work demonstrated BEC of polaritons [5,6] and magnons [7].

Classical systems can also show condensation effects. We theoretically and experimentally found, for example that the light in actively mode-locked lasers shows in some cases condensation behavior [8,9]. In the condensate state an optimum short pulse, which corresponds to the lowest pulse mode, becomes dominant in the cavity. It was also noted there [8] that in higher than 1-dimensional laser mode systems transverse spatial loss modulation or spectral filtering can lead to condensation. We also mention other important theoretical work on condensation phenomena with classical nonlinear waves [10], disordered lasers [11] and the review in Ref. [12]. The many-body nature of laser modes has been discussed in earlier work. We developed a broad thermodynamic-like approach for mode-locked lasers showing, for example, that mode-locking is a phase transition [13–17], and it also exhibits critical phenomena [18,19], in the deep statistical mechanics meaning, where noise takes the role of temperature.

In the present work we deal with a generic mode system with noise, represented by a basic multi-mode cw laser cavity, without nonlinear terms, that nevertheless is found to exhibit a route to condensation which is formally very similar to BEC in a potential-well [20]. This laser light condensation (LC) phenomenon is based on very different grounds than regular BEC. Its "energy" levels are measured in a loss-gain scale, inherent in laser cavities, where the "ground-state" is the lowest-loss mode, and noise, also inherent in lasers, has the role of temperature. The loss scale gives a mode occupation hierarchy and power spectra that resemble the Bose-Einstein distribution and leads to laser light-condensation (LC). It has special properties compared to regular BEC, as we discuss below. For example, the condensed state can be anywhere in the spectral band, in contrast to BEC where it is always at the lowest energy (frequency) state.

It would be therefore interesting to raise the possibility that photon-BEC experiments in optical cavities [3] can fall in a classical LC or lasing category rather than being a quantum-thermal based photon-BEC phenomenon. We shall argue that LC and photon condensation do not provide a new type of photons ("super-photon"?) or a new "quantum" light state, not more than a single-frequency laser does. It is however a challenging topic that needs further study and experimental work.

# 2. Mathematical route to laser light condensation (LC)

The ingredients for LC to happen are very simple while having a generic nature: many modes ("particles"), noise (spontaneous emission, thermal etc.) which injects light into those modes, a gain part, and certain conditions on the loss-gain filtering spectrum (loss "potential-well") as discussed below. A typical example for such system is a cw laser. The equations of motion for the evolution of light modes circulating in the cavity are taken in a most simple form, without dispersion, modulation and nonlinear terms [21], or other linear or nonlinear rate equations [22,23], but with a noise source [13–19]. We have a multivariate Langevin equation (with shared and variate-dependent coefficients, shown to be a key part for condensation):

$$\frac{da_m}{d\tau} = (g - \varepsilon_m)a_m + \Gamma_m, \qquad (1)$$

Where $a_m(\tau)$ are the slowly varying electric field complex amplitudes of the modes (in $d$-dimensions: $a_m \to a_{m_1...m_d}$), $\tau$ is the long term time variable that counts cavity roundtrip frames, and $g$ is a slow saturable gain factor, shared by all modes. It functions as a Lagrange multiplier (and $\mu \equiv g - \varepsilon_0$ as "chemical potential") for setting the overall cavity power $P$, and $\varepsilon_m$ is spectral filtering (loss dispersion) due to frequency and mode dependent loss, absorption and gain. $\Gamma_m$ is an additive noise term that can originate from spontaneous emission, or any other internal or external sources. It is modeled by a white Gaussian process with covariance $2T$:

$$<\Gamma_m(\tau)\Gamma_n^*(\tau')> = 2T\delta(\tau-\tau')\delta_{mn}, \text{ and } <\Gamma_m(\tau)> = 0,$$

where $<>$ denotes average. We describe the modes frequency by $\Omega = \omega - \omega_0 = (c/n)(|\bar{k}| - k_0)$ (where $\bar{k}$ is the wavevector and $n$ is the refractive index), measured with respect to the linewidth center $\omega_0 = (ck_0/n)$. The modes discrete relative frequencies are $\Omega_m$. With the continuous variable we have $a_m(\tau) \to a(\Omega,\tau)$, and $\varepsilon_m \to \varepsilon(\Omega)$. We note that our study is applicable to discrete modes, as well as to continuous frequencies, alluding to the common dual (but not necessarily the same) terminology and meaning of single-mode and single-frequency lasers.

An important part that we add in this work, compared to the usual laser formulation [22,23], is to allow and examine various spectral dependences for $\varepsilon(\Omega)$, that is usually taken to be parabolic. This is a key point that opens the way for condensation. We show in this work that for certain spectral filtering functions $\varepsilon(\Omega)$, a linear mode system with noise provides a route to condensation with a transition of noise broadened spectra to a single-frequency oscillation.

We turn to the analysis that leads to LC, which resembles the BEC formalism in a potential-well [20]. We obtain from Eq. (1), at steady state, the overall power:

$$P = \sum_m p_m = \sum_m <a_m a^*_m> = \sum_m \frac{T}{\varepsilon_m - g} \;\to\; \frac{T}{\varepsilon_0 - g} + T\int_0^{\varepsilon_N} \frac{\rho(\varepsilon)d\varepsilon}{\varepsilon + \varepsilon_0 - g} \qquad (2)$$

For the integral at the right hand side of the equation with the continuous net loss variable, we define the density of loss states $\rho(\varepsilon)$ that is shown to have a prime role for condensation occurrence. The first term at the right hand side of Eq. (2) gives the power of the lowest loss ($\varepsilon_0$) mode, and the second term is the power in all of the higher modes. We can right away notice the resemblance to BEC in a potential-well [20]. The weight for each spectral component depends on $\rho(\varepsilon)$ and a factor $1/(\varepsilon + \varepsilon_0 - g)$ that replaces the Bose-Einstein statistics. Here the upper limit of the integral, $\varepsilon_N$, is set by the spectral filtering (loss "potential-well") depth. However, it is the density of the low loss-levels at $\varepsilon \approx 0$ that determines the integral convergence. We therefore take the spectral filtering functional dependence as a power law around $\varepsilon_0$: $\varepsilon = \gamma |\Omega|^\eta = \tilde{\gamma}|\tilde{\Omega}|^\eta$ where $\tilde{\Omega} = \Omega/\Omega_0$, $\tilde{\gamma} = \gamma \Omega_0^\eta$ and $\Omega_0$ is the spectral filtering width. It can be shown that it gives the following density of states (modes) in $d$-dimensions [9,20]:

$$\rho(\varepsilon) = \frac{V_d}{(2\pi)^d} \int d^d k \, \delta(\varepsilon - \gamma |\Omega|^\eta) \simeq \left(\frac{C_d}{\eta \gamma^{\xi+1}}\right) \varepsilon^\xi , \qquad (3)$$

where $C_d = V_d b_d \, n k_0^{d-1}/[(2\pi)^d c]$, $\xi = \eta^{-1} - 1$, $V_d$ is the $d$-dimensional resonator volume, $\delta(x)$ is the Dirac delta function, and $b_d = 2, 2\pi, 4\pi$ for $d = 1, 2, 3$, respectively. We could also use for the calculation $\rho(\varepsilon) = (V_d/(2\pi)^d) \int dS / |\bar{\nabla}_k \varepsilon|$, where $dS$ is a surface element at constant $\varepsilon$ in the $\bar{k}$ space.

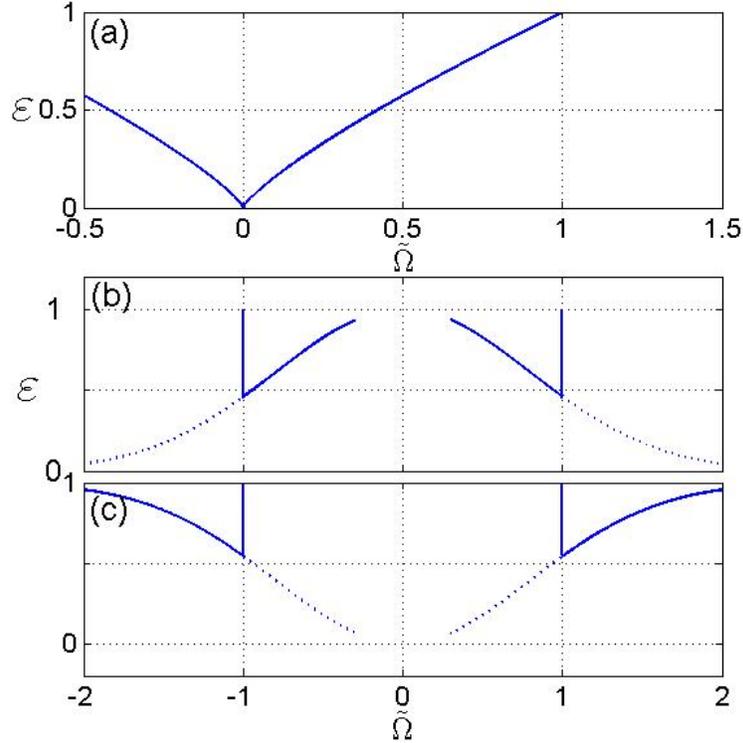

Fig. 1. Spectral filtering: (a) Power law function with $\tilde{\gamma} = 1$ and $\eta = 0.8$. (b), (c) Cut structures (solid line) obtained for example from sections of Gaussian (can be exponential etc.) filtering spectra (combined frequency dependent losses/absorption and gain). They can result from various frequency cutoff mechanisms, such as the far apart longitudinal mode separation in microcavity lasers [3,4]. The cuts lowest loss ($\varepsilon_0$) mode at $\tilde{\Omega} = \pm 1$ are redefined to be at $\tilde{\Omega} = 0$.

We now return to Eq. (2), and specifically to the integral at the right hand side. Condensation occurs when the integral converges at $\varepsilon = 0$. Then the power population of the light at higher than $\varepsilon = 0$ levels stays unchanged (filled) at $P_c = T \int_0^{\varepsilon_N} \rho(\varepsilon)d\varepsilon/\varepsilon$. Therefore, additional pumping that increases $P$ beyond $P_c$ must be channeled into the lowest level power $p_0 = T/(\varepsilon_0 - g)$ which starts growing macroscopically. The integral converges when $\xi > 0$, i.e. $\eta < 1$ (nonsmooth $\varepsilon(\Omega)$ at $\Omega = 0$ and concave, at least at $\Omega = \pm 0$, as shown in the illustrative examples in Fig. 1). At condensation ($P \geq P_c$), the net gain of the lowest mode ("chemical potential) is $\mu \equiv g - \varepsilon_0 = 0$.

We note that $\varepsilon(\Omega)$ doesn't have to be symmetric for condensation as long as the slope of $\varepsilon(\Omega)$ is smaller than 1 at $\Omega = 0$. Examples are the spectral slices in Fig. 1 cut from broad spectra (e.g., Gaussian, exponential) that can result from various reasons, such as boundary-condition caused cutoffs. We also note that except for the prefactor in $\rho(\varepsilon)$ the condensation condition doesn't depend on the dimension *d*, except when additional spatial or temporal filtering or modulation exists on other dimensions (e.g. transverse) as noted in Ref. [8].

The noise induced spectrum (mode-population) is given by the integrand of Eq. (2) expressed with the $\tilde{\Omega} = \Omega/\Omega_0$ variable:

$$p(\tilde{\Omega}) = \frac{C_d T}{\varepsilon(\tilde{\Omega}) + \varepsilon_0 - g} \ . \tag{4}$$

For the power law spectral filtering, we have $p(\tilde{\Omega}) = C_d T /[\tilde{\gamma}\tilde{\Omega}^\eta + \varepsilon_0 - g]$. Figure 2a shows spectra for $\eta = 0, 8$ and various overall power values $P$ (or $\mu$ s), below and at condensation.

More generally, we can have other spectral filtering functions. Then, since the condensation is determined by the exponent near $\Omega = 0$, it is required to have for condensation: $\left.\frac{d\varepsilon}{d\Omega}\right|_{\Omega=\pm 0} < 1$. Away from $\Omega = 0$, $\varepsilon(\Omega)$ can have any dependence that experiments yield. In many cases there are exponential sections that can be expressed around $\varepsilon_0$ as $\varepsilon(\Omega) = \gamma_e(e^{\beta|\Omega|} - 1)$. Here the slope at $\Omega = \pm 0$ is $\eta = 1$, just on the verge of condensation. We can then obtain for the mode density, $\rho(\varepsilon) = C_d/[\beta(\varepsilon + \gamma_e)]$, and for the spectrum (beyond $\Omega = 0$):

$$p(\tilde{\Omega}) = \frac{C_d T}{\gamma_e(e^{\tilde{\beta}|\tilde{\Omega}|} - 1) + (\varepsilon_0 - g)}, \tag{5}$$

where $\tilde{\beta} = \Omega_0 \beta$. This equation with the exponential term resembles thermal Bose-Einstein distribution and has implications to the discussion below on the LC connection with photon-BEC experiments. Figure 2b shows graphs of this spectrum for various $\mu$ s. We can see the broad thermal-like exponential dependence.

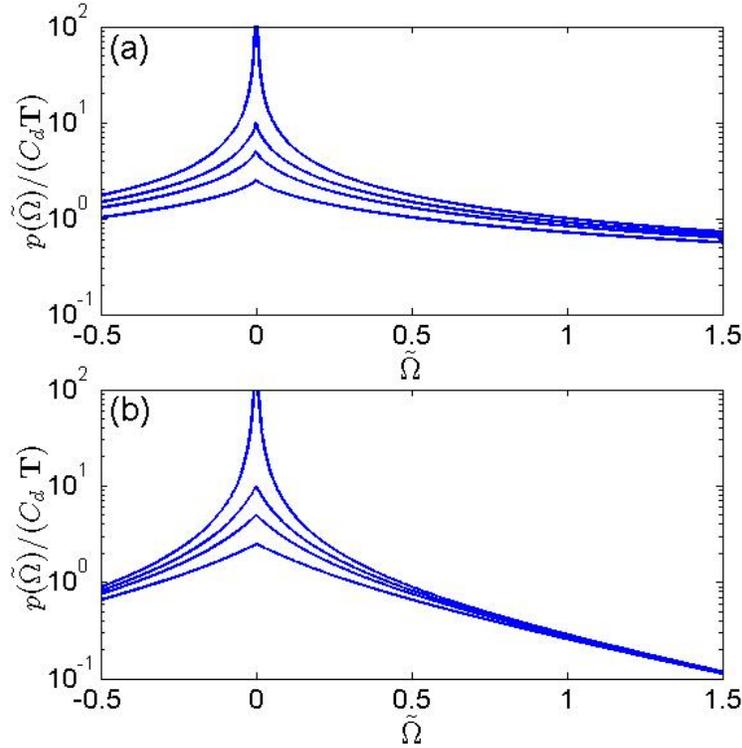

Fig. 2. Light spectra, in a semi-logarithmic scale, obtained by Eqs. (4) and 5: (a) For the power law dependence spectral filtering example (Fig. 1a), with $\tilde{\gamma} = 1$ and $\eta = 0.8$. The various graphs correspond (from bottom to top) to $\mu \equiv g - \varepsilon_0 = -0.4, -0.2, -0.1, 0$, and respectively to $P/P_c = 0.23, 0.3, 0.4, 1$. In the cut examples (the high and low pass filters in Fig. 1b) the spectra have only the right or left of $\Omega = 0$ sides. (b) For the exponential spectral filtering given by Eq. (5), with the above $\mu$s, $\gamma_e = 1$ and $\tilde{\beta} = 1.5$. In both figures, the three lower graphs correspond to non-condensate states, and the upper one to condensation. At and beyond the condensation transition, $\mu = 0$, and the spectrum stays as in the top graph, except for the lowest-loss mode $p_0$ at $\Omega = 0$ that grows upon further pumping (not shown here, but given in Fig. 4).

Figure 3 shows the "chemical potential" ($\mu \equiv g - \varepsilon_0$) dependence on $P$, for $\eta = 0.8$. $\mu$ is always negative before condensation, and becomes zero at condensation. The condensation is shown in Fig. 4 by the power dependence of the normalized low-loss mode $p_0/P_c$. Above the phase transition ($P \geq P_c$) all the excess power goes to the condensation state $p_0$. This is the laser LC route to condensation, characterized by the sharp transition from a multi- to single-mode oscillation.

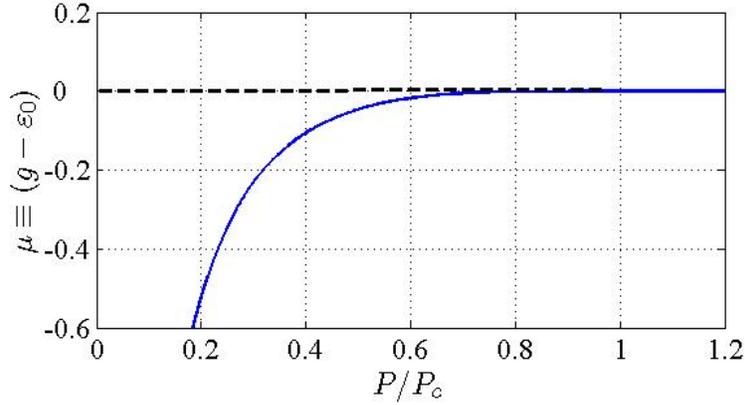

Fig. 3. Normalized "chemical potential" $\mu = (g - \varepsilon_0)$ as a function of the overall normalized laser power $P/P_c$, for $\tilde{\gamma}=1$, $\eta = 0.8$, and $\varepsilon_N = 1$.

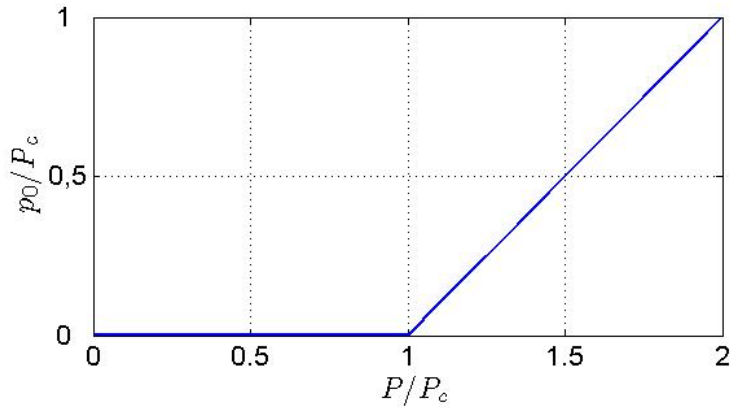

Fig. 4. Condensation transition: Normalized lowest mode power $p_0/P_c$ as a function of the overall normalized laser power for for $\tilde{\gamma}=1$, $\eta = 0.8$, and $\varepsilon_N = 1$. Above the phase transition ($P \geq P_c$) all the excess power goes to $p_0$ - the condensation state.

## 3. Discussion on lasing, LC, photon-BEC and experiments

LC is formally similar to BEC in having:

   a. Non-interacting particles - bosons in BEC and light-modes in LC.
   b. Particle distribution - Bose-Einstein statistics vs. loss-dependent mode weighting in LC.
   c. A global constraint on the overall bosonic particles number in BEC vs. the overall light-modes power in LC.

  d. The chemical potential in BEC corresponds to the lowest-mode gain minus loss in LC. In both cases it is negative below condensation, and approaches and becomes zero at and above the transition.

  e. Certain conditions on the density of states function for bosonic particle and light-modes.

  f. A similar mathematical route to condensation in BEC and LC.

There are however a few important differences between LC and BEC, besides the basic one that LC is classical and BEC is a quantum-thermal effect. We specify the main ones:

  a. The states (modes) hierarchy ("energy ruler") in LC is measured by loss-gain, compared to photon energy (frequency) in BEC.

  b. It is a general noisy system in LC vs. thermal equilibrium, characterized by temperature, in BEC.

  c. Laser LC occurs at the lowest-loss state ($\Omega = 0$), which can be anywhere in the spectral band, and not necessarily at the lowest photon energy ($\hbar\omega$) state, as it is in BEC.

  d. The spectrum of the oscillating modes in LC (Fig. 2) is dictated by the loss scale. It is "noisization" rather than the BEC thermalization spectrum [4]. Unlike BEC, the spectra can span over frequencies above, or below, or at both sides of $\Omega = 0$. We also note that in LC, below $P_c$, there is a hole in $p(\varepsilon)$ near $\Omega = 0$, ($|\bar{k}| = k_0$), resulting from $\rho(\Omega = 0) = 0$, but not in $p(\Omega)$ (in the frequency domain), as seen in Fig. 2. In BEC, prior to condensation, there is a hole in the energy distribution at zero energy (as in black-body radiation spectra), which is always at the lowest possible energy (frequency) [20].

  e. Unlike regular BEC [20], there is no dimensional dependence in LC when the condensation results from spectral filtering, that operates in one dimension ($\Omega$, at the $|\bar{k}| \approx k_0$ sphere shell), except for a factor that results from the d-dimensional sphere surface. Nevertheless, when there are in the system other space (or time) dependent loss-gain filtering (or modulation), like in two dimensional transverse mode systems [3,4], the condition for condensation can be different [8].

We elaborate on the LC and BEC connection and the relation to lasers and experiments. Both cases are characterized by a transition to a single-frequency oscillation. Such a transition can also occur in regular lasers, like in ideally homogeneous lasers [23]. In a way, every single-frequency laser could have been regarded as a condensate state. (Photon densities in single-frequency lasers can be much higher than in the microcavity experiment [3].) In those lasers the single-mode oscillation is explained by a nonlinear saturation effect or a simple ad hoc assumption that the first mode to reach threshold clamps the gain profile and therefore eliminates the oscillation of all other modes [23]. In LC, it is embedded in the density of states where high mode states populated by the noise become filled (for certain quasi-continuous mode densities $\rho(\varepsilon)$ with exponents $\eta < 1$), leading to condensation, without an explicit nonlinearity, but with a global constraint, in a similar way to BEC. It is likewise characterized by a negative "chemical potential" $\mu \equiv g - \varepsilon_0$ that becomes zero at condensation (as shown in Fig. 3.) It means that we have a different route to a single-mode oscillation than regular lasers.

In experiments it can be very challenging to relate and identify the different single-frequency oscillation effects, since they are all done in laser cavities. It is possible that experiments in laser cavities fall in LC or a single-mode lasing category, rather than being a thermal photon-BEC phenomenon [3]. We have shown the role of spectral-filtering, inevitable in lasers, that governs the photon gas "climate", the "equilibrium" and the spectra in a pumped many-mode cavity with gain. In the dye-filled microcavity experiment [3,4], it includes the transverse-modes loss-hierarchy (mode-filtering) due to the difference (even if very weak) in their transverse waveform, mirror-reflectivity, gain, absorption and other losses. The frequency cutoffs due to far apart longitudinal modes in the microcavity can give the needed abrupt spectral filtering profile. However, even without spectral filtering, LC can occur in the lateral domain. The situation can be similar to what we suggested in Ref. [8] (see there the last paragraph before conclusion), where lateral modulation or confinement, and transverse mode filtering, even if the modes had the same frequency, can cause transversal condensation [8]. Then the condition on the exponent of the lateral loss profile ("potential-well") is very much relaxed. In two-dimensions, any finite positive $\eta$ near $\Omega = 0$ gives condensation.

There are also other issues, discussed elsewhere and briefly given here, that need consideration in the microcavity experiments [3]. The massively lost photon replacement in the cavity by pumping occurs via spectral-loss criterion. It is generally the case in lasers, being non-Hamiltonian with frequency dependent dissipation, not in thermal equilibrium, and not a grand-canonical ensemble. The microcavity is not an exception. The seemingly large photon "keepers" (the mirrors) do not provide better performance than what we usually have in many other lasers. The microcavity mirrors have ultra high reflectivity of 0.999985 [3,4], and therefore provide a very high finesse. Nevertheless, since the cavity length is small (~1.5µm), important parameters value stay in the usual regime. The photon loss through the mirrors by itself gives a cavity photon lifetime of ~0.3*ns*, that is similar or shorter than usual values in lasers. Likewise, the gain coefficient needed for steady state oscillation is relatively large (~0.2/*cm*). When adding other losses, absorption and scatterings, one obtains even much shorter cavity photon lifetimes and higher gain coefficients. It means that the photon population in the cavity is changing very rapidly. The photon number (power) is kept by the pumping that induces stimulated emission photons which replace in steady state the lost photons.

Broadened mode-spectra resulting from (amplified) spontaneous-emission (is random phase noise) and stimulated-emission, such as those obtained with the transverse-modes in the dye-filled microcavity experiment [3,4], basically exist without thermal excitations, at any temperature. Such spectra will not "cool-down" to a single frequency by lowering the temperature, say to ~0°*K*, if it were not a single-mode lasing but a thermal-BEC effect. (Therefore, it would be crucial to measure in experiments like Ref. 3 the temperature dependence of the spectrum in a broad temperature range, and not only at or close to room temperatures.) There can be thermal effect in the spectra, since the dye molecules [3,4] alone can have within their sub-levels some thermal distribution (with questions and limitations [24]) which affects the photon emission. The microcavity experiment however is not a free dye-molecules system. Therefore the spectral filtering is not only the dye molecules absorption-emission part upon which the Kennard-Stephanov (KS) relation [24] is based. Cavities impose on the photons their own additional rules.

The collapse to a single frequency at the red side of the band in the micro-cavity experiment [3] is supposed to support the photon-BEC view. BEC dictates condensation at the lowest possible frequency (the extreme "red" side. Therefore, if there is further availability in the spectrum at the red side (of the "cut-off"), the BEC peak had to move to there. It is therefore important to confirm that the spectrum cannot extend to even lower frequencies. In the microcavity experiment [3,4], it seems that there can be a potential spectral region of transverse modes below the semi "cut-off". (The short cavity makes the longitudinal modes far apart but doesn't fully cut the spectrum). It is the region of the high transverse modes of the next (to the "red" side) longitudinal mode. Therefore, in a BEC view, the condensation frequency has to move to even lower frequencies. Otherwise, one has to talk about isolated "spectral islands" free of other than thermal constraints, where photons might have regional thermal equilibrium. This possibility would need however further study and verification. In the lasing LC view however it can be simple. The idea there is that the loss-scale, and not the frequency as it is in BEC, determines the light spectrum in a laser cavity. The red side of the cut-off has higher losses (high transverse modes of the next red longitudinal mode) and so the light there is damped. It therefore supports the laser LC approach for the microcavity experiment. The spectrum with the cut-off due to the far apart longitudinal modes provides the needed abrupt loss spectral-filtering profile for the LC phenomenon. Moreover, we have already argued that even much less restrictive condition on the filtering leads to laser condensation in two-dimensional laser systems.

LC can thus explain the microcavity experimental results [3], including the spectra below and above the collapse to a single frequency oscillation, associating the low mode population near the frequency semi-cutoff as lasing or LC at the lowest-loss transverse mode, and not as photon BEC. We also stress that LC can produce and explain exponentially dependent spectra, similar to the thermal behavior that was attributed to Bose-Einstein distribution [3,4], as we have shown above (Eq. (5) and Fig. 2b) for common exponential loss (absorption and gain) spectral filtering functions.

We mention again that photon densities in single-frequency lasers can be as high as, and higher, than what was calculated for a photon-BEC realization, where it was said that "photon wave packets spatially overlap" [3]. One may raise questions about that, and argue if there is any new (quantum?) meaning there and in other regular ("classical"?) high photon density systems, such as LC and single-frequency lasers.

We conclude this discussion by noting that although we raised questions on experiments on photon-BEC, we don't exclude the possibility to observe it. We think however that the whole issue would need further discussion and study, especially of properties that differentiate BEC from single-frequency lasing and LC.

Realization of LC can be obtained in many-mode laser systems. In particular, fiber lasers (one dimensional), such as erbium-doped lasers, are very suitable since they can have many and dense longitudinal modes in their gain bandwidth. For the loss spectrum, it is possible to fabricate synthesized fiber gratings with reflection frequency profiles needed for condensation. So can be used other lasers such as the 2-dimensional microcavity [3], relying on the transverse modes. In this case, LC condensation can result from either spectral filtering or mode filtering and spatial loss modulation [8]. Abrupt spectra can be produced by the frequency semi-cutoffs of far apart longitudinal modes in microcavities, while mode filtering and spatial modulation by lateral confinement. LC is then expected to occur in transverse mode systems [3], as was discussed above.

## 4. Conclusion

We have presented a classical laser light condensation phenomenon (LC) which is based on a loss hierarchy in a linear mode system with noise and certain spectral filtering. We discussed experimental sides, differences between LC and BEC and difficulties to observe regular photon-BEC in laser cavities. Further experimental study is needed to verify all those issues.

We finally note that the formalism that leads to LC, based on Eq. (1), can have a broad scope. It presents a generic many "particle" Langevin equation with friction coefficients composed of an independent dispersive part $\varepsilon_m$, and a shared part ($g$). Besides lasers, we can think of various mechanical or biological particles in fluids which follow this model and can therefore show condensation behavior.

### Acknowledgments

This research was supported by the US-Israel Binational Science Foundation (BSF) and the Israel Science Foundation (ISF). I thank Alexander Bekker and Oded Fischer for valuable help.